\documentclass[mypaper,7pt,twoside]{CoAst}
\usepackage{epsf,graphicx,fancyhdr}
\renewcommand{\chaptermark}[1]{\markboth{#1}{}}

\newcommand{\kopf}{\small\itshape Comm. in Asteroseismology \\ Contribution to the Proceedings of the Wroclaw HELAS Workshop, 2008}

\newcommand{\Authors}[1]{\begin{center}\normalsize\bf\sf #1 \end{center}}

\renewcommand{\author}[1]{\begin{center}\normalsize\bf\sf #1 \end{center}}
\newcommand{\Address}[1]{\begin{center}\small\sf #1 \end{center}}

\newcommand{\Session}[1]{{\vspace{3mm}\small \noindent  \hspace*{3mm} Session: } #1 \normalsize}

\newcommand{\Objects}[1]{{\vspace{0mm}\small \noindent  \hspace*{3mm} Individual Objects: } \small #1 \normalsize}

	\newcommand{\threeB}{\small STARS - opacity driving, levitation, opacity data \newline}

\renewenvironment{abstract}{\section*{Abstract}\normalsize\sf}{}
\newcommand{\References}[1]{\begin{flushleft}{\large References\\}\vspace*{2mm}\small #1 \end{flushleft}}

\newcommand{\chapterCoAst}[2]{\chapter[\sf\normalsize #1\\ \footnotesize \hspace*{5mm}by #2 \sf\normalsize][]{#1\\}\rhead[\fancyplain{}{\sf\footnotesize \center{#1}}]{\fancyplain{}{\sffamily\thepage}}\lhead[\fancyplain{\kopf}{\sffamily\thepage}]{\fancyplain{\kopf}{\sf\footnotesize \center{#2}}}}

\newcommand{\figureCoAst}[5]{\begin{figure}[#4]
\centering
\includegraphics*[#5]{#1}
\caption{#2}
\label{#3}
\end{figure}}

\renewcommand{\chaptername}{}

\newcommand{\acknowledgments}[1]{\vspace*{5mm}\noindent  \textbf{Acknowledgments.} #1}

\def\rfr{\smallskip\par\noindent
        \hangindent=7truemm
        \hangafter=1}

\begin{document}
\sf

\chapterCoAst{Long-term photometric monitoring of the hybrid subdwarf B pulsator HS\,0702+6043}
{R.\,Lutz, S.\,Schuh, R.\,Silvotti, et al.} 
\Authors{R.\,Lutz$^{1,2}$, S.\,Schuh$^{1}$, R.\,Silvotti$^3$, R.\,Kruspe$^1$,
  and S.\,Dreizler$^1$}
\Address{
$^1$ Institut f\"ur Astrophysik, Friedrich-Hund-Platz 1, 37077 G\"ottingen, Germany\\
$^2$ Max-Planck-Institut f\"ur Sonnensystemforschung, Max-Planck-Stra\ss e 2,\\
  37191 Katlenburg-Lindau, Germany\\
$^3$ INAF - Osservatorio Astronomico di Capodimonte, via Moiariello 16,\\80131
  Napoli, Italy
}

\noindent
\begin{abstract}
Pulsating subdwarf B stars oscillate in short-period $p$-modes or
long-period $g$-modes. HS\,0702+6043 is one of currently three
objects known to show characteristics of both types and hence is
classified as hybrid pulsator. We briefly present our analysis of the
$g$-mode domain of this star, but focus on first results from
long-term photometric monitoring in particular of the $p$-mode
oscillations. We present a high-resolution frequency spectrum,
and report on our efforts to construct a multi-season O--C diagram.
Additionally to the standard (although nontrivial) exercise in
asteroseismology to probe the instantaneous inner structure of a star,
measured changes in the pulsation frequencies as derived from an
\mbox{O--C} diagram can be compared to theoretical evolutionary timescales.
Within the \mbox{EXOTIME} program, we also use this same data to
search for planetary companions around extreme horizontal branch objects.
\end{abstract}

\Session{ \threeB }
\Objects{HS\,0702+6043, HS\,2201+2610}

\section*{Introduction}
Subdwarf B stars (sdB stars) populate the extreme horizontal
branch (EHB) at effective temperatures between 20\,000 and 40\,000~K
and surface gravities $\log (g/\rm{cm\,s^{-2}})$ between 5.0 and
6.2. They are believed to be core helium-burning objects of half a solar
mass with remaining hydrogen envelopes too thin to sustain H-shell burning.
The reason for losing almost all of their original hydrogen envelope in
earlier stages of their evolution is still unknown.
The high fraction of binaries among the sdB stars
suggests that close binary evolution may play an important role in
their formation. The discovery of a planetary companion to
HS\,2201+2610 (Silvotti et al.\ 2007),
the only case where the necessary kind
of measurements for such a discovery are available so far for an sdB star,
now revives the idea that planets in wide orbits may also play a
role in the formation of these stars.\\
A fraction of the sdB stars shows pulsations.
Variable subdwarf B stars (sdBV stars) can be divided into the classes
of rapid \emph{$p$-mode pulsators} (sdBV$_\textrm{r}$) and slow
\emph{$g$-mode pulsators} (sdBV$_\textrm{s}$),
with three objects known so far to belong to both classes simultaneously
(\emph{hybrid pulsators}, sdBV$_\textrm{rs}$). These are of particular
interest since the two mode types probe different regions within the star.
Both types of pulsations are driven by a
$\kappa$-mechanism where the required opacity bump is due to iron and
nickel accumulated by diffusion, resulting in phase-stable pulsational
behaviour (Charpinet et al.\ 1997; Fontaine et al.\ 2003;
Jeffery \& Saio 2006).\\
The $p$-mode pulsators show low amplitudes (few ten mmag) and short periods
(few minutes) at higher temperatures (roughly 30\,000-35\,000~K). In contrast,
the $g$-mode pulsators have even lower amplitudes (few mmag) and longer
periods (30 to 90~min) at lower temperatures (roughly 25\,000-30\,000~K).
In a $\log g$\,--\,$T_{\rm eff}$ diagram, the hybrids are
located at the interface of the $p$-mode and $g$-mode instability regions
(see Figure 1 in Lutz et al.\ 2008a).
The class prototypes are EC\,14026-2647 (Kilkenny et al.\ 1997)
and PG\,1716+426 (Green et al.\ 2003), respectively. The known
hybrids are HS\,0702+6043 (Schuh et al.\ 2006),
Balloon\,090100001 (Oreiro et al.\ 2005; Baran et al.\ 2005)
and HS\,2201+2610 (Lutz et al.\ 2008b).
Asteroseismology has been one of the important tools to
constrain the evolutionary history of subdwarf B stars.
While an asteroseismological solution provides an instantaneous
''snapshot'' of the interior stellar structure, extended monitoring
of \emph{changes} over several years in the short, stable $p$-mode
pulsation frequencies allows to directly measure evolutionary
timescales $\dot{P}$ which can be compared to predictions
from evolutionary models. $\dot{P}$ can be measured from
O--C diagrams, which will at the same time reveal the presence of
potential planets that may in turn have influenced the previous evolution.

\section*{EXOTIME}
EXOTIME is the abbreviation of EXOplanet search with the TIming MEthod. This
program is led by Roberto Silvotti and Sonja Schuh. Being a collaborative
long-term campaign, it involves a substantial number of observers and telescopes all
over the world which perform ground based time-series \mbox{photometry}.
Basic information (target objects, observing schedule etc.\ ) can be found on
the program's webpage
\mbox{http://www.na.astro.it/$\sim$silvotti/exotime/}.
While most of the more than 300 exoplanets are found around main sequence host
stars, \mbox{EXOTIME} searches planets orbiting evolved sdB pulsators, i.e.\ extreme
horizontal branch objects. Currently the target list consists of five
pulsating sdB stars, HS\,0702+6043 and HS\,2201+2610 being two of them. The
search for exoplanets is performed with the timing method or O--C analysis,
which also allows to derive evolutionary timescales (described on the example
of HS\,2201+2610 in a following section). Closely related to the
search for \mbox{exoplanets} orbiting sdB stars are evolutionary aspects of sdB
stars, late-stage or post RG evolution of planetary systems and the question
if planets could be responsible for the extreme mass loss of sdB progenitors.

\section*{HS\,0702+6043}
The sdB pulsator HS\,0702+6043 was first identified as a variable in a search
program by Dreizler et al.\ (2002). Its spectroscopic parameters place it at
the common boundary of the $p$- and $g$-mode instability regions and as
mentioned above, Schuh et al.\ (2006) indeed first revealed this object to be
a hybrid pulsator. This detection triggered extensive follow-up observations
which are listed in Table 1.
\begin{table}[!t]
\caption{Photometric data of HS\,0702+6043. CA: Calar Alto, T: T\"ubingen, SB:
Steward Bok, G: G\"ottingen, L: Loiano, MB: Mt.Bigelow.}
\begin{center}
\begin{tabular}{lrr|lrr}
\hline
Date & Site & Length & Date & Site & Length \\
\hline
Dec 1999 & CA 1.2m & 8.4h & Feb 2008 & T 0.8m & 20.1h\\
Feb 2004 & T 0.8m & 7.3h & Feb 2008 & G 0.5m & 32.2h\\
Feb 2004 & SB 2.2m & 12.0h & Mar 2008 & G 0.5m & 0.6h\\
Jan 2005 & CA 2.2m & 56.0h & Mar 2008 & L 1.5m & 8.0h\\
Dec 2007 & T 0.8m & 31.8h & Nov07 - Mar08 & MB 1.55m & 424.0h\\
Dec 2007 & G 0.5m & 12.0h & May 2008 & G 0.5m & 4.1h\\
\hline
\end{tabular}
\end{center}
\end{table}
There are photometric data available going back to 1999, but unfortunately
with large gaps in between. A regular monitoring was not performed until the
end of 2007.

\subsection*{$p$-modes}
\figureCoAst{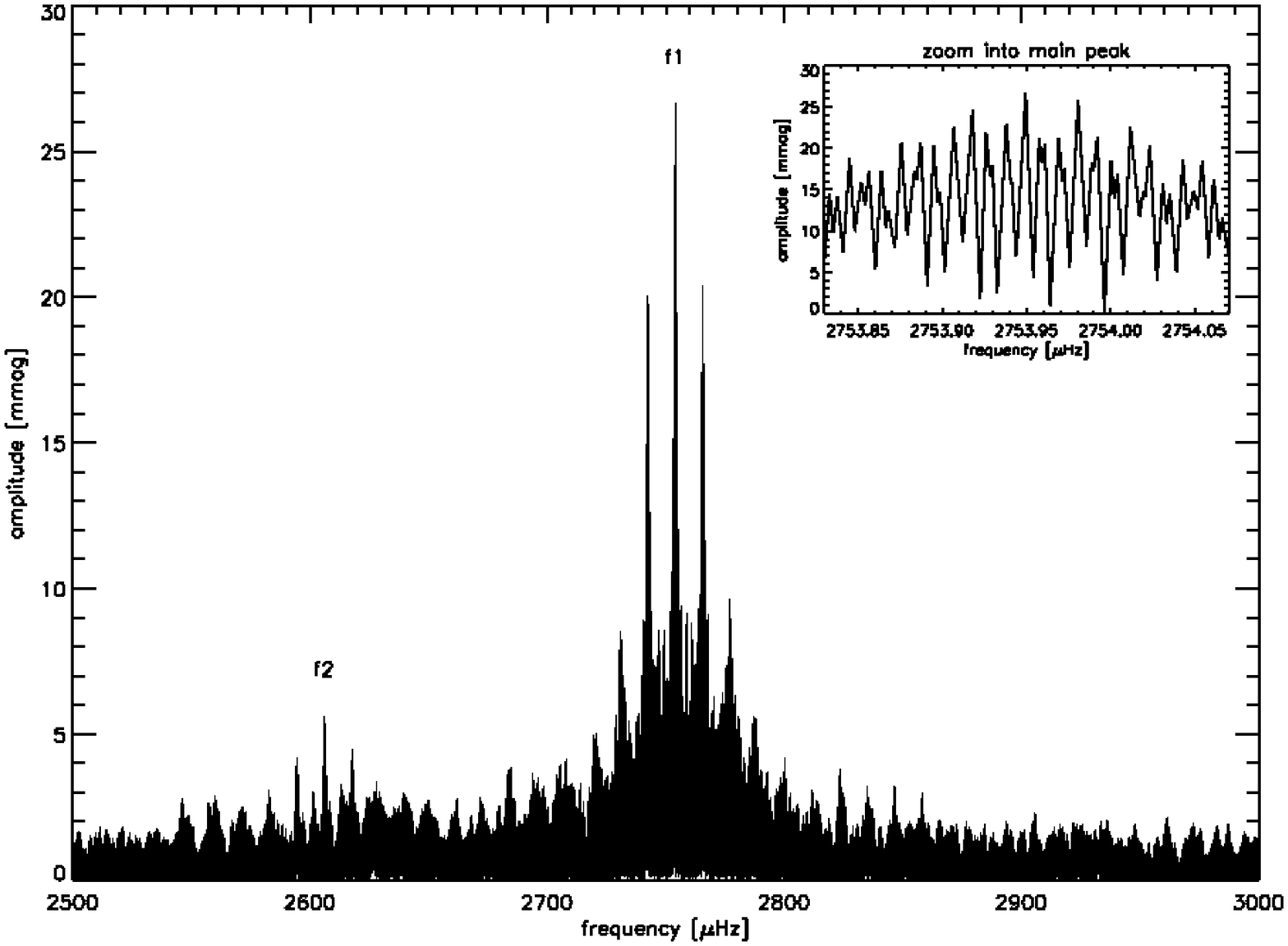}{Periodogram of HS\,0702+6043 showing the two main
  pulsation periods f1 and f2. The inset window is a zoom into the main-peak.}
{pmodes}{!ht}{clip,angle=0,width=115mm}

The short-period $p$-modes are the relevant ones for the construction of a
multi-seasonal O--C diagram. From all the data until Feb 2008 (MB not
included), we derived the high-resolution frequency spectrum displayed in
Figure \ref{pmodes}, which shows the two dominant frequency features f1 and f2 at
frequencies of 2753.9\,$\mu$Hz (363.1\,s) and 2606.1\,$\mu$Hz (383.7\,s),
respectively. The corresponding amplitudes are 26.6 and 5.5\,mmag. Due to the
large gaps in our data archive, we cannot present a meaningful O--C analysis
for HS\,0702+6043 yet, but with a regular EXOTIME monitoring we are confident
that this will be possible within the next two years.

\subsection*{$g$-modes}
The Jan 2005 run (see Table 1) was initiated in order to resolve the low
frequency $g$-mode regime in HS\,0702+6043 and indeed, three features could be
identified in the data at frequencies of 271.7\,$\mu$Hz, 318.1\,$\mu$Hz and
206.3\,$\mu$Hz (61.3\,min, 52.4\,min, 80.8\,min) with amplitudes of 1.8\,mmag,
1.3\,mmag and 0.9\,mmag, respectively. A periodogram can be found in
Lutz et al.\ (2008a).

\figureCoAst{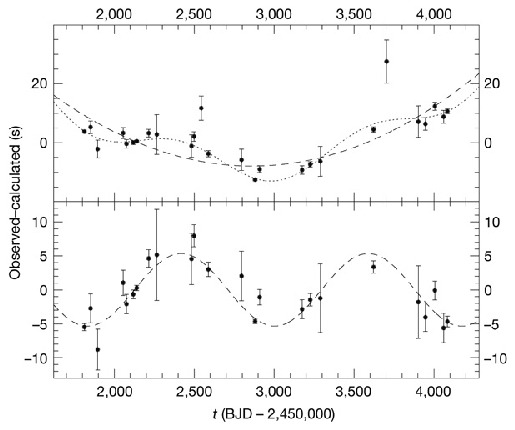}{O--C diagram of the main pulsation
  frequency of HS\,2201+2610. Each dot represents a whole campaign. Top panel:
  parabolic fit, bottom panel: sinusoidal residuals. The sinusoidal component
  has a semi-amplitude of 5.3\,s. Plot by R.Silvotti.}{o-c}{!ht}
  {clip,angle=0,width=115mm}

\section*{HS\,2201+2610}
HS\,2201+2610 is so far the only EXOTIME target for which an extended O--C
analysis could be executed (Silvotti et al.\ 2007). Regular data sets are
available since 2000.

\subsection*{O--C Analysis}
An O--C (Observed minus Calculated) analysis is a way to measure the phase
variations of a periodic function. The observed times of the pulsation maxima
(or minima) of the single runs of an observational season are compared to the
calculated ephemeris of the whole data set (see e.g.\ Kepler et al.\ 1991).
Different shapes in the O--C can be identified with different situations:
a parabolic shape indicates a linearly changing period, whereas a sinusoidal
component can be interpreted by cyclically advanced and delayed timings of the
pulsation maxima (or minima) due to the presence of a low-mass companion, which
causes the pulsator to wobble around the common barycenter.
Figure \ref{o-c} shows the O--C analysis for the main pulsation period of
HS\,2201+2610, which is around 350\,s at an amplitude of about 10\,mmag.
The parabolic shape of the O--C diagram in the top panel of Figure \ref{o-c}
indicates a linearly changing period of $\dot{P}=1.46\cdot 10^{-12}$. Since
this is a positive value, one can infer that this object is expanding,
i.e.\ cooling. The evolutionary timescale can be calculated as
$P/\dot{P}=7.6$\,Myr, consistent with theoretical predictions
(Charpinet et al.\ 2002).
The bottom panel of Figure \ref{o-c} displays the sinusoidal residuals which
are induced by the gravitational influence of a planetary mass body. Some
system parameters that can be derived are the companion's mass of 3.2 Jupiter
masses (still with an uncertainty due to the unknown inclination), an orbital
period of 1170 days, an orbital separation of 1.7\,AU, a star projected
orbital velocity of 99\,m/s or a planet orbital velocity of 16\,km/s.
For more system parameter and a detailed description of the applied
assumptions refer to Silvotti et al.\ (2007).

\acknowledgments{
RL thanks the organizers for financial support. The authors thank all
observers who contributed observations to the HS\,0702+6043 and
HS\,2201+2610 data archive.
}

\References{
\rfr Baran, A., Pigulski, A., Kozie{\l}, D., et al.\ 2005, MNRAS, 360, 737
\rfr Charpinet, S., Fontaine, G., Brassard, P., et al.\ 1997, ApJ, 483, L123
\rfr Charpinet, S., Fontaine, G., Brassard, P., \& Dorman, B.\ 2002, ApJ, 140, 469
\rfr Dreizler, S., Schuh, S., Deetjen, J.L., et al.\ 2002, A\&A, 386, 249
\rfr Fontaine, G., Brassard, P., Charpinet, S., et al.\ 2003, ApJ, 597, 518
\rfr Green, E.M., Fontaine, G., Reed, M.D., et al.\ 2003, ApJ, 583, L31
\rfr Jeffery, C.S., \& Saio, H.\ 2006, MNRAS, 372, L48
\rfr Kepler, S.O., Winget, D.E., Nather, R.E., et al.\ 1991, ApJ, 378, 45
\rfr Kilkenny, D., Koen, C., O'Donoghue, D., \& Stobie, R.S.\ 1997, MNRAS,
285, 640
\rfr Lutz, R., Schuh, S., Silvotti, R., et al.\ 2008a, ASPC, 392, 339
\rfr Lutz, R., Schuh, S., Silvotti, R., et al.\ 2008b, A\&A, submitted
\rfr Oreiro, R., P{\'e}rez Hern{\'a}ndez, F., Ulla, A., et al.\ 2005, A\&A,
438, 257
\rfr Schuh, S., Huber, J., Dreizler, S., et al.\ 2006, A\&A, 445, L31
\rfr Silvotti, R., Schuh, S., Janulis, R., et al.\ 2007, Nature, 449, 189
}

\end{document}